\documentclass[aps,prd,nofootinbib,groupedaddress]{revtex4}
\usepackage{graphicx}
\usepackage{epsfig}
\usepackage{dcolumn}
\usepackage{amsfonts}
\usepackage{amssymb}
\usepackage{bm}
\usepackage{amsmath}
\usepackage{pdfpages}
\usepackage{color}
\def\Journal#1#2#3#4{{#1} {\bf#2}, #3 (#4)}

\def\NPA{{\rm Nucl. Phys.} A}
\def\NPB{{\rm Nucl. Phys.} B}
\def\PLB{{\rm Phys. Lett.}  B}
\def\PRL{\rm Phys. Rev. Lett.}
\def\PRD{{\rm Phys. Rev.} D}
\def\PRC{{\rm Phys. Rev.} C}

\def\EPJC{{\rm Eur. Phys. J.} C}

\def\la{\langle}
\def\ra{\rangle}

\def\lam{\lambda}
\def\al{\alpha}

\def\be{\begin{equation}}
\def\ee{\end{equation}}
\def\bea{\begin{eqnarray}}
\def\eea{\end{eqnarray}}
\begin{document}
\title{Two-particle twist-3 distribution amplitudes of the pion and kaon in the light-front quark model}
\author{ Ho-Meoyng Choi\\
{\em Department of Physics, Teachers College, Kyungpook National University,
     Daegu, Korea 41566}\\
         Chueng-Ryong Ji\\
{\em Department of Physics, North Carolina State University,
Raleigh, NC 27695-8202} }
\begin{abstract}
We investigate the two-particle twist-3 distribution amplitudes (DAs) of the pseudoscalar mesons, in particular pseudoscalar ($\phi^P_{3;M}(x)$) and 
pseudotensor ($\phi^\sigma_{3;M}(x)$) DAs of pion and kaon, 
in the light-front quark model based on the variational principle.
We find that the behavior of the conformal symmetry in each meson distribution amplitude depends on the chiral limit characteristics of the 
light-front trial wave function taken in the variational principle.
We specifically take the two different light-front trial wave functions, Gaussian vs. power-law type, and discuss their characteristics of the conformal symmetry
in the chiral symmetry limit as well as their resulting degree of the conformal symmetry breaking in $\phi^P_{3;M}(x)$ and $\phi^\sigma_{3;M}(x)$
depending on the trial wave function taken in the computation. We present numerical results of transverse moments, Gegenbauer-moments and $\xi$-moments
and compare them with other available model estimates. 
The SU(3) flavor-symmetry breaking effect is also quantified with the numerical computation.  
\end{abstract}

\maketitle
\section{Introduction}
\label{sec:I}
Hadronic distribution amplitudes (DAs) are the longitudinal projection of the hadronic wave functions obtained by integrating the transverse
momenta of the fundamental constituents~\cite{BL80,ER80,CZ84}.
These nonperturbative quantities are defined as vacuum-to-hadron matrix elements
of particular nonlocal quark or  quark-gluon operators and
thus encode important information on bound states in QCD.
Especially, the electromagnetic and transition form factors at high $Q^2$ as well as the $B$-physics phenomenology in the context of SU(3) flavor symmetry breaking effect require a detailed information of meson DAs. 
Meson DAs are also indispensable for the analysis of hard exclusive electroproduction based on the QCD factorization~\cite{Collins-Frankfurt-Strikman}. In particular, the shape of the pion DA has been extensively discussed due to the nature of the pion as the massless Nambu-Goldstone boson~\cite{Radyushkin,Polyakov}.
Finding the fundamental nonperturbative information of QCD
motivated many theoretical studies to calculate meson DAs using nonperturbative methods
such as the QCD sum rule~\cite{CZ84,BF,Ball99,BBL,HWZ04,HZW05,Agaev,BMS06,MPS10,SP15}, the chiral-quark
model from the instanton vacuum~\cite{Goeke,NK06,NKHM06},
the Nambu-Jona-Lasinio (NJL) model~\cite{AB02,PR}, the Dyson-Schwinger
equation (DSE) approach~\cite{DSE13,DSE2015}, and the light-front quark model~(LFQM)~\cite{CJ_DA,Hwang10}.
Among them, the LFQM appears to be one of the most effective and efficient tools in studying hadron physics as it takes advantage of the distinguished features of the light-front dynamics (LFD)~\cite{BPP}. Working in Minkowski space, the LFD allows the study of physical observables both in spacelike and timelike kinematic regions.
The rational energy-momentum dispersion relation  of LFD, namely
$p^-=({\bf p}^2_\perp + m^2) / p^+$, yields the sign correlation between the light-front~(LF) energy $p^-(=p^0-p^3)$ and the LF longitudinal momentum $p^+(=p^0 + p^3)$ and leads to the suppression of vacuum fluctuations in LFD. It facilitates the partonic interpretation of the hadronic amplitudes. 
The LFD also carries the maximum number~ (seven) of the kinetic (or interaction independent) generators and 
thus the less effort in dynamics is necessary in order to get the QCD solutions that reflect the full Poincar$\acute{e}$ symmetries.
Based on the advantage of LFD, the LFQM
has been quite successful in describing various static and non-static
properties of hadrons~\cite{CJ_99,CJLR15,Choi07,Jaus90,Jaus91,schlumpf,
Cheng97,CCP,Card95,MFPS02,CJK02,CJ_kaon,CJ06,CJ08,CJ_Bc,CJ_GPD01,CYA}
such as meson mass spectra~\cite{CJ_99,CJLR15}, the decay constants (i.e. the lowest moments of light-cone DAs)~\cite{Choi07,Hwang10},
electromagnetic and weak transition form factors~\cite{Jaus90,Jaus91,schlumpf,Cheng97,CCP,Card95,MFPS02,CJK02,CJ_kaon,CJ06,CJ08,CJ_Bc} and
generalized parton distributions~(GPDs)~\cite{CJ_GPD01,CYA}.
The LFQM analysis of the pion form factor~\cite{CJ06,CJ08} has also provided compatible results
both in spacelike and timelike regions with the holographic approach to LF QCD~\cite{Brodsky2008}
based on the 5-dimensional anti-de Sitter (AdS) spacetime and the conformal
symmetry which has given insight into the nature of the effective confinement potential and the resulting LF wave functions for both
light and heavy mesons~\cite{Brodsky2006}.

Through the recent analysis of the twist-2 and twist-3 DAs of pseudoscalar and vector
mesons~\cite{CJ2015,FB2015,CJ_V14,LC2016}, 
we discussed also the link between the chiral symmetry of QCD and the LFQM. 
In Ref.~\cite{CJ_DA}, we have analyzed the two-particle twist-2
DAs of pseudoscalar~($\phi^A_{2;M}(x)$) and vector~($\phi^{||}_{2;V}(x)$) mesons
using our LFQM~\cite{CJ_99}. We then extended our LFQM to analyze
two-particle twist-3 pseudoscalar ($\phi^P_{3;M}(x)$) DAs of
pseudoscalar mesons~\cite{CJ2015,FB2015} and chirality-even
twist-3 ($\phi^{\perp}_{3;V}(x)$) DAs of vector mesons~\cite{CJ_V14} to
discuss the link between the chiral symmetry of QCD and the numerical results of the LFQM.
In particular, through the analysis of twist-3 DAs of $\pi$ and $\rho$ mesons, we observed that the LFQM with effective degrees of freedom represented by the constituent quark and antiquark could provide the view of effective zero-mode cloud around the quark and antiquark inside the meson. 
Our numerical results appeared consistent with this view and effectively indicated that the constituent quark and antiquark in the LFQM 
could be considered as the dressed constituents including the zero-mode quantum fluctuations from the vacuum.

To discuss the wave function dependence of the LF
zero-mode~\cite{Zero1,Zero2,Zero3,Jaus99,BCJ01,BCJ03,Cheng04,CJ_PV,MF12,Tobias}
contributions to $\phi^P_{3;M}(x)$ and $\phi^{\perp}_{3;V}(x)$,
we analyzed both the exactly solvable manifestly covariant Bethe-Salpeter (BS) model and 
the more phenomenologically accessible realistic LFQM~\cite{CJ_99,CJ_DA} in the standard LF
approach. The purpose of taking the exactly solvable covariant BS model was to check the existence (or absence)
of the zero mode in each channel, e.g. $\phi^P_{3;M}(x)$ or $\phi^{\perp}_{3;V}(x)$, without any ambiguity.
For example, performing the LF calculation in the covariant BS model with the multipole type $q{\bar q}$ bound state
vertex function, we not only showed that the twist-3 $\phi^P_{3;M}(x)$ and $\phi^{\perp}_{3;V}(x)$
receive both the zero-mode and the instantaneous contributions
but also identified the zero-mode operator corresponding to the zero-mode contribution. 
As discussed in Refs.~\cite{CJ_V14,CJ2015}, we also found the
universal mapping [see e.g. Eq. (35) in~\cite{CJ2015}] between the covariant BS model and 
the standard LFQM
for any two-point and three-point functions.  With this mapping, we were able to boost the exactly solvable 
covariant BS model computation into the more phenomenologically accessible LFQM computation. 
In practice, the LF vertex function obtained in the covariant BS model was mapped into the phenomenological, 
typically Gaussian, LF 
trial wave function 
which has been scrutinized by the standard LFQM analysis of meson mass spectroscopy based on the variational principle and other meson phenomenology~\cite{CJ_99,CJLR15}. 
The remarkable finding from this practice was that the zero-mode
contribution as well as the instantaneous contribution
revealed in the covariant BS model became absent in the
LFQM with the LF on-mass-shell constituent quark and antiquark degrees of freedom. Without involving the zero-mode
and instantaneous contributions, our LFQM with the Gaussian trial wave function provided the result 
of twist-3 DAs $\phi^P_{3;M}(x)$ and $\phi^{\perp}_{3;V}(x)$ which not only satisfied the fundamental
constraint (i.e., symmetric form with respect to $x$) anticipated
from the isospin symmetry but also provided the consistency both with the chiral symmetry and the conformal symmetry
(e.g., the correct asymptotic form in the $m_q\to 0$ limit) expected from the QCD. Our LFQM predictions with the Gaussian wave function such as
$\phi^{||}_{2;\rho}(x)\to 6x(1-x)$ and $\phi^{\perp}_{3;\rho}(x)\to (3/4)[1+(2x-1)^2]$ for $\rho$
and $\phi^P_{3;\pi}(x)\to 1$ for $\pi$
in the chiral symmetry limit reproduce the exact functional forms anticipated from QCD's
conformal limit~\cite{Ball98,BF}.
This exemplifies that our LFQM prediction with the Gaussian wave function satisfies both the chiral symmetry and the conformal symmetry consistent
with the QCD if one correctly implements the zero-mode link to the QCD vacuum.

It is important, however, to realize that satisfying both the chiral symmetry and the conformal symmetry depends on the choice of the LF trial wave function.
The key in the Gaussian LF wave function is the factorization of the transverse momentum dependence from the dependence of scale independent parameters such as mass. 
It allows that the $m_q\to 0$ limit satisfies both the chiral symmetry and the conformal symmetry simultaneously.
If the LF trial wave function is not taken as Gaussian but for example taken as 
power-law (PL) type, 
then 
the factorization of the transverse momentum dependence from the scale independent parameter dependence cannot be fulfilled 
and thus the $m_q\to 0$ limit may not satisfy the conformal symmetry although it may still satisfy the chiral symmetry. 
This dependence on the LF trial wave function indicates that some particular meson DAs may not satisfy the conformal symmetry
while they still satisfy the chiral symmetry consistent with QCD. 
Similarly, the DSE approach in~\cite{DSE2015} provided the asymptotic form
of the pion $\phi^P_{3;\pi}(x)$ with a broad downward concave shape in the central region of $x$ rather than $\phi^P_{3;\pi}(x)\to 1$ anticipated
from  QCD's conformal limit~\cite{BF}.
There are two independent two-particle twist-3 DAs of a pseudoscalar meson, namely, pseudoscalar DA $\phi^P_{3;M}$ and pseudotensor DA $\phi^\sigma_{3;M}$~\cite{BF,Ball99,BBL,PR,NK06,HWZ04,HZW05}.
The authors in~\cite{DSE2015} also analyzed the pseudotensor DA $\phi^\sigma_{3;M}(x)$, and found that the asymptotic form of the pion $\phi^\sigma_{3;\pi}(x)$ coincide with the anticipated expression of QCD's conformal limit, $6 x(1-x)$.

These developments motivate our present work for the more-in-depth analysis of the two-particle twist-3 pion and kaon DAs in LFQM 
with different forms of LF trial wave functions.
We first extend our previous work~\cite{CJ2015} to analyze the twist-3
pseudotensor DA $\phi^\sigma_{3;M}(x)$ of a pseudoscalar meson within the LFQM. We also
discuss the discrepancy of the asymptotic forms of $\phi^P_{3;\pi}(x)$ between
DSE approach~\cite{DSE2015} and QCD's conformal limit expression~\cite{BF} from the perspective of
dependence of DA on the form of LF trial wave functions such as
Gaussian wave function vs. PL wave function.
Although the two-particle twist-3 pion DAs were briefly discussed in 
LC2016~\cite{LC2016}, 
we elaborate more in this work on the dependence
of DA on the form of LF trial wave functions as well as the SU(3) flavor-symmetry breaking effect through the complete analysis of 
two-particle twist-3 DAs of pseudoscalar meson.
In order to compute the twist-3 pseudotensor DA $\phi^\sigma_{3;M}(x)$,
we again utilize the same manifestly covariant BS model used in~\cite{CJ2015,FB2015,CJ_V14} 
to check the existence (or absence) of the LF zero-mode contribution. We then apply 
the previously found universal mapping [see e.g. Eq. (35) in~\cite{CJ2015}] between the covariant BS model and 
the standard LFQM to map the vertex function obtained in the exactly solvable covariant BS model into the more phenomenologically accessible 
Gaussian and PL radial wave functions provided from our LFQM variational principle computation.

The paper is organized as follows.
In Sec.~\ref{sec:II}, 
we compute the twist-3 pseudotensor DA $\phi^\sigma_{3;M}(x)$ in an exactly solvable model based on the covariant BS model of (3+1)-dimensional fermion field theory.
We then link the covariant BS model to the standard LFQM with the previously found universal mapping
between the two as discussed above and present the resulting form of $\phi^\sigma_{3;M}(x)$ as well as $\phi^P_{3;M}(x)$ in our
LFQM. In Sec.~\ref{sec:III}, we present our numerical results of
$\phi^\sigma_{3;M}(x)$ and $\phi^P_{3;M}(x)$ for the pion and kaon and discuss the results
in the chiral vs. conformal symmetry limit. The SU(3) flavor symmetry breaking effects on the twist-3 DAs
for the kaon are also discussed. Summary and discussion follow in Sec.~\ref{sec:IV}. In the Appendix,
the derivation of twist-3 DAs of pseudoscalar meson is presented.

\section{Model Description}
\label{sec:II}
\subsection{Manifestly Covariant BS Model}

\begin{figure}
\begin{center}
\includegraphics[height=3.5cm, width=7cm]{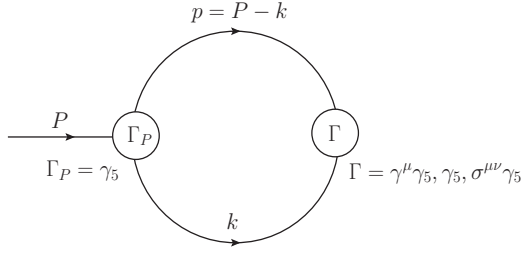}
\caption{\label{fig1} Feynman diagram for the one-quark-loop evaluation of the meson
decay amplitude in the momentum space. }
\end{center}
\end{figure}
The $\phi^P_{3;M}$ and $\phi^\sigma_{3;M}$ are defined in terms of the
following matrix elements of gauge invariant nonlocal operators
in the light-front gauge~\cite{BF,Ball99,BBL}:
\be\label{Deq:1}
\la 0|{\bar q}(z)i\gamma_5 q(-z)|M(P)\ra
= f_M \mu_M \int^1_0 dx e^{i\zeta P\cdot z} \phi^P_{3;M}(x),
\ee
and
\be\label{Deq:2}
\la 0|{\bar q}(z)\sigma_{\alpha\beta}\gamma_5 q(-z)|M(P)\ra
=-\frac{i}{3} f_M \mu_M (P_\alpha z_\beta - P_\beta z_\alpha)
\int^1_0 dx e^{i\zeta P\cdot z} \phi^\sigma_{3;M}(x),
\ee
where $z^2=0$
and $P$ is the four-momentum of the meson ($P^2=m^2_M$) and the integration variable $x$ corresponds to the longitudinal momentum fraction
carried by the quark and $\zeta =2x -1$ for the short-hand notation.
The normalization parameter $\mu_M = m^2_M /(m_q + m_{\bar q})$
results from quark condensate. For the pion, $\mu_\pi = -2\la {\bar q}q\ra / f^2_\pi$ from the Gell-Mann-Oakes-Renner relation~\cite{GOR}.
We also note from the argument in~\cite{DSE13,DSE2015} that the pseudoscalar DA of the pion,
$\phi^P_{3;\pi}(x)$, i.e. pseudoscalar projection of the pion's LF wave function,
might be understood as describing the probability distribution of the chiral condensate within the
pion~\cite{BRST}.
The normalization of the two twist-3 DAs $\Phi=\{ \phi^P_{3;M}, \phi^\sigma_{3;M} \}$ is given by
\be\label{Deq:3}
\int^1_0 dx \; \Phi(x) = 1.
\ee

Defining $z^\mu = \tau \eta^\mu$ using the lightlike vector $\eta=(1,0,0,-1)$, one can rewrite Eqs.~(\ref{Deq:1}) and~(\ref{Deq:2}) as [see Appendix for the explicit derivation
of Eqs.~(\ref{Deq:3.1}) and~(\ref{Deq:3.2})]
\be\label{Deq:3.1}
\phi^P_{3;M}(x) = \frac{2 (P\cdot\eta)}{f_M \mu_M}
\int^\infty_{-\infty}  \frac{d\tau}{2\pi} e^{-i\zeta\tau(P\cdot\eta)} \la 0|{\bar q}(\tau \eta)i\gamma_5 q(-\tau\eta)|M(P)\ra,
\ee
and
\be\label{Deq:3.2}
\phi^\sigma_{3;M}(x) = - \frac{12}{f_M \mu_M}
\int^\infty_{-\infty}  \frac{d\tau}{2\pi} \int^x_0 dx'
e^{-i\zeta'\tau(P\cdot\eta)}
\la 0|{\bar q}(\tau\eta) i(\slash\!\!\!\!P \slash\!\!\!\eta - P\cdot\eta)\gamma_5 q(-\tau\eta)|M(P)\ra,
\ee
respectively.  The nonlocal matrix elements
${\cal M}_{\alpha} \equiv \la 0|{\bar q}(\tau\eta) i\Gamma_\alpha q(-\tau\eta) |M(P)\ra$
for pseudoscalar ($\Gamma_\alpha=\gamma_5$) and pseudotensor ($\Gamma_\alpha=(\slash\!\!\!\!P \slash\!\!\!\eta - P\cdot\eta)\gamma_5$) channels are
given by the following momentum integral in two-point function of the manifestly covariant BS model (see Fig.~\ref{fig1})
\be\label{Deq:4}
{\cal M}_\alpha = N_c
\int\frac{d^4k}{(2\pi)^4} e^{-i \tau k\cdot\eta} e^{-i \tau(k-P)\cdot\eta}
\frac{H_0} {N_p N_k} S_\alpha,
\ee
where $N_c$ denotes the number of colors and $S_\alpha =  {\rm Tr}\left[i\gamma_5\left(\slash \!\!\!p+m_q \right)
 \gamma_5 \left(-\slash \!\!\!k + m_{\bar q} \right) \right]$ for pseudoscalar channel and
 ${\rm Tr}\left[i(\slash\!\!\!\!P \slash\!\!\!\eta - P\cdot\eta)\gamma_5\left(\slash \!\!\!p+m_q \right)
 \gamma_5 \left(-\slash \!\!\!k + m_{\bar q} \right) \right]$ for pseudotensor channel.
The denominators $N_p (= p^2 -m^2_q +i\varepsilon)$
and $N_k(= k^2 - m^2_{\bar q}+i\varepsilon)$ come from the quark propagators
of mass $m_q$ and $m_{\bar q}$ carrying the internal four-momenta $p =P -k$ and $k$, respectively.
In order to regularize the covariant loop,
we use the usual multipole ansatz~\cite{Jaus99,CJ_V14,MF97,SS} for the $q{\bar q}$ bound-state vertex function
$H_0=H_0(p^2,k^2)$ of a meson:
$H_0(p^2,k^2) = g/N_\Lambda^n$,
where $N_\Lambda  = p^2 - \Lambda^2 +i\varepsilon$, and $g$ and $\Lambda$ are constant parameters.
We note that the power $n$ for the multipole ansatz should be $n\geq 2$ to regularize the loop integral
and our essential results in terms of the zero-mode issue do not depend on the value of $n$.

For the LF calculation, we use the metric convention 
$a\cdot b =\frac{1}{2}(a^+b^- + a^-b^+)-{\bf a}_\perp\cdot{\bf b}_{\perp}$
and separate the trace term $S_\al$ into the on-mass-shell propagating part
$[S_\al]_{\rm on}$ and the off-mass-shell instantaneous part $[S_\al]_{\rm inst}$,
i.e. $S_\al = [S_\al]_{\rm on} + [S_\al]_{\rm inst}$
 via 
$\slash\!\!\!q=\slash\!\!\! q_{\rm on} + \frac{1}{2}\gamma^+(q^- - q^-_{\rm on})$.
In the reference frame where ${\bf P}_\perp =0$, i.e.,
$P=( P^+, M^2/P^+, 0)$, the LF energies of the on-mass-shell
quark and antiquark are given by
$ p^-_{\rm on} = ({\bf k}^2_\perp + m^2_q)/ xP^+$ and
$ k^-_{\rm on} = ({\bf k}^2_\perp + m^2_{\bar q})/(1-x) P^+$, respectively,
where $x=p^+/P^+$ is the LF longitudinal momentum fraction of the quark.

After a little manipulation, we can rewrite Eq.~(\ref{Deq:3.1}) for the pseudoscalar channel as
\bea\label{Deq:5.1}
\phi^P_{3;M}(x) &=& \frac{N_c}{f_M\mu_M}
\int\frac{d^4k}{(2\pi)^4} \delta \biggl(1 - x - \frac{k\cdot\eta}{P\cdot\eta} \biggr)
\frac{H_0} {N_p N_k} S_P
\nonumber\\
&=&  \frac{N_c}{f_M\mu_M} \int
 \frac{d^2{\bf k}_\perp}{16\pi^3}
 \frac{\chi(x,{\bf k}_\perp)}{(1-x)} [S_P]_{\rm full},
\eea
where 
\be\label{Deq:6}
\chi(x,{\bf k}_\perp) = \frac{g}{[x (m_M^2 -M^2_0)][x (m_M^2 - M^2_{\Lambda})]^n},
\ee
and
\be
\label{Deq:6-1}
 M^2_{0(\Lambda)} = \frac{{\bf k}^{2}_\perp + m^2_q(\Lambda^2)}{x}
 + \frac{ {\bf k}^{2}_\perp + m^2_{\bar q}}{1-x}.
\ee
The full result of the trace term $[S_P]_{\rm full}$ has been obtained in~\cite{CJ2015}
and it receives not only $[S_P]_{\rm on}$ and $[S_P]_{\rm inst}$ but also
the zero-mode contribution $[S_P]_{\rm Z.M.}$ in this manifestly covariant BS model,
i.e. $[S_P]_{\rm full} = [S_P]_{\rm on} + [S_P]_{\rm inst} +[S_P]_{\rm Z.M.}$, where
$[S_P]_{\rm on}=4 ( p_{\rm on}\cdot k_{\rm on} + m_q m_{\bar q} )= 2 [ M^2_0 - (m_q - m_{\bar q})^2]$,
$[S_P]_{\rm inst}=2k^+ (p^- - p^-_{\rm on}) = 2 (1-x) (m^2_M - M^2_0)$, and 
$[S_P]_{\rm Z.M.}=-2[x (m^2_M -M^2_0) + m^2_q - m^2_{\bar q} + (1-2x)m^2_M]$,
respectively. The detailed procedure to obtain the zero-mode calculation is given in~\cite{CJ2015}.
However, as we have explained in great detail in~\cite{CJ2015}, the full result of trace term $[S_P]_{\rm full}$ in the more
realistic LFQM using the Gaussian or PL type wave functions gives the same result for the decay amplitude
only with the on-mass-shell contribution involving neither the zero-mode contribution nor the instantaneous 
contribution. Effectively, it indicates that the on-mass-shell constituent quark and antiquark in the LFQM 
can be considered as the dressed constituents including the zero-mode and instantaneous quantum fluctuations from the vacuum.
The same observation has been made for the calculation of the twist-2 and-3 DAs of the vector meson~\cite{CJ_V14}
as well as the pion electromagnetic form factor~\cite{CJ2015}.

Similarly, Eq.~(\ref{Deq:3.2}) for the pseudotensor twist-3 $\phi^\sigma_{3;M}(x)$ can be rewritten as
\bea\label{Deq:5.2}
\phi^\sigma_{3;M}(x) &=& -\frac{6}{f_M\mu_M}\frac{N_c}{(P\cdot\eta)}
\int\frac{d^4k}{(2\pi)^4} \int^x_0 dx' \delta \biggl(1 - x' - \frac{k \cdot \eta}{P \cdot \eta} \biggr)
\frac{H_0} {N_p N_k} S_\sigma,
\nonumber\\
&=& -\frac{6}{f_M\mu_M}\frac{N_c}{P^+} \int
 \frac{d^2{\bf k}_\perp}{16\pi^3} \int^x_0 dx'
 \frac{\chi(x',{\bf k}_\perp)}{(1-x')} [S_\sigma]_{\rm full},
\eea
where $\chi(x',{\bf k}_\perp)=\chi(x\to x',{\bf k}_\perp)$. We should note for this pseudotensor channel that,
due to the nature of the second rank tensor operator contracting meson momentum, the r.h.s. of Eq.~(A.6) is not the DA itself but the derivative of DA so that
the $x'$-integration appears in Eq.~(\ref{Deq:5.2}) with the integration range from 0 to $x$. 
One may find without any difficulty that the manifestly covariant calculation of the trace term $S_\sigma$ would give zero result for the decay amplitude
if the $x'$-integration is done from zero to 1 since DA at the end point $x=1$ must be zero. 
As the $x'$ integration range from 0 to $x$, the decay amplitude is in general not zero unless $x=1$ or $x=0$.
In the LF calculation, the same observation can be made
if we include all three contributions, i.e. on-mass-shell, instantaneous, and zero-mode contributions, in the full result of the trace term 
$[S_\sigma]_{\rm full} = [S_\sigma]_{\rm on} + [S_\sigma]_{\rm inst} +[S_\sigma]_{\rm Z.M.}$, where
$[S_\sigma]_{\rm on}=4[ (P\cdot k_{\rm on}) p^+ - (P\cdot p_{\rm on}) k^+] = 2 P^+ [ (2x'-1) M'^2_0 + m^2_{\bar q} - m^2_q]$,
$[S_\sigma]_{\rm inst}=-2k^+ P^+ (p^- - p^-_{\rm on}) = -2 P^+ (1-x') (m^2_M - M'^2_0)$, and 
$[S_\sigma]_{\rm Z.M.}=2P^+[x' (m^2_M -M'^2_0) + m^2_q - m^2_{\bar q} + (1-2x')m^2_M]$ with $M'_0 = M_0 (x\to x')$,
respectively. This indicates that not only the on-mass-shell contribution but also both the instantaneous contribution and the zero-mode
contribution in principle exist in the LF calculation to coincide with the manifestly covariant BS result.  
 However, it is remarkable to observe that the full result of trace term $[S_\sigma]_{\rm full}$ in the more
realistic LFQM using the Gaussian or PL type wave functions which we discuss in the next subsection, Sec.\ref{sec:IIIb}, is identical to the result 
when $[S_\sigma]_{\rm full}$ is replaced 
by $[S_\sigma]_{\rm on}$ as discussed in the case of pseudoscalar channel. It assures that the on-mass-shell constituent quark and antiquark
in the LFQM can be regarded as the dressed constituents including the zero-mode and instantaneous quantum fluctuations from the vacuum.
\subsection{Application to Standard Light-Front Quark Model}
\label{sec:IIIb}
In the standard LFQM~\cite{CJ_99,CJLR15,Choi07,Jaus90,Jaus91,schlumpf,
Cheng97,CCP,Card95,MFPS02,CJK02,CJ_kaon,CJ06,CJ08,CJ_Bc,CJ_GPD01,CYA}, the
wave function of a ground state pseudoscalar meson ($J^{\rm PC}=0^{-+}$)
as a $q\bar{q}$ bound state is given by
\be\label{QM1}
\Psi_{\lam{\bar\lam}}(x,{\bf k}_{\perp})
={\Phi_R(x,{\bf k}_{\perp})\cal R}_{\lam{\bar\lam}}(x,{\bf k}_{\perp}),
\ee
where $\Phi_R$ is the radial wave function and the
spin-orbit wave function ${\cal R}_{\lam{\bar\lam}}$
with the helicity $\lam({\bar\lam})$ of a quark(antiquark)
that is obtained by the interaction-independent Melosh transformation~\cite{Melosh}
from the ordinary spin-orbit wave function assigned by the quantum numbers $J^{PC}$.
The covariant form of the spin-orbit wave function ${\cal R}_{\lam{\bar\lam}}$
is given by
\be\label{QM4}
{\cal R}_{\lam{\bar\lam}}
=\frac{\bar{u}_{\lam}(p_q)\gamma_5 v_{{\bar\lam}}( p_{\bar q})}
{\sqrt{2}[M^{2}_{0}-(m_q -m_{\bar q})^{2}]^{1/2}},
\ee
and it satisfies
$\sum_{\lam{\bar\lam}}{\cal R}_{\lam{\bar\lam}}^{\dagger}{\cal R}_{\lam{\bar\lam}}=1$.
The normalization of our wave function is then given by
\be\label{QM6}
\sum_{\lam{\bar\lam}}\int\frac{dx d^2{\bf k}_\perp}{16\pi^3}
|\Psi_{\lam{\bar\lam}}(x,{\bf k}_{\perp})|^2
=
\int\frac{dx d^2{\bf k}_\perp}{16\pi^3}
|\Phi_R(x,{\bf k}_{\perp})|^2.
\ee

For the radial wave function $\Phi_R$,
we try both the Gaussian or harmonic oscillator (HO) wave function $\Phi_{\rm HO}$ and the power-law (PL) type
wave function $\Phi_{\rm PL}$~\cite{schlumpf} as follows
\be\label{QM2}
\Phi_{\rm HO}(x,{\bf k}_{\perp})=
\frac{4\pi^{3/4}}{\beta^{3/2}} \sqrt{\frac{\partial
k_z}{\partial x}} {\rm exp}(-{\vec k}^2/2\beta^2),
\ee
and
\be\label{QM2_PL}
\Phi_{\rm PL}(x,{\bf k}_{\perp})= \sqrt{\frac{128\pi}{\beta^3}}
\sqrt{\frac{\partial k_z}{\partial x}}
\frac{1}{(1 + {\vec k}^2/\beta^2)^2},
\ee
where $\vec{k}^2={\bf k}^2_\perp + k^2_z$ and $\beta$ is the variational parameter
fixed by the analysis of meson mass spectra~\cite{CJ_99}.
The longitudinal component $k_z$ is defined by $k_z=(x-1/2)M_0 +
(m^2_{\bar q}-m^2_q)/2M_0$, and the Jacobian of the variable transformation
$\{x,{\bf k}_\perp\}\to {\vec k}=({\bf k}_\perp, k_z)$ is given by
\be\label{QM3}
\frac{\partial k_z}{\partial x}
= \frac{M_0}{4 x (1-x)} \biggl\{ 1-
\biggl[\frac{m^2_q - m^2_{\bar q}}{M^2_0}\biggr]^2\biggr\}.
\ee
As discussed in the previous section, Sec. \ref{sec:I}, the transverse momentum ${\bf k}_{\perp}$ dependence factorizes
as ${\rm exp}(-{\vec k}^2/2\beta^2) = {\rm exp}(-{\bf k}^2_\perp/2\beta^2) {\rm exp}(-k^2_z/2\beta^2)$ in $\Phi_{\rm HO}$
while such factorization of ${\bf k}_{\perp}$ dependence of ${1}/{(1 + {\vec k}^2/\beta^2)^2}$ is not feasible in $\Phi_{\rm PL}$.
Thus, the scale (or conformal) invariance of the transverse momentum ${\bf k}_{\perp}$ as well as the longitudinal momentum fraction $x$ is
achieved in the massless (chiral) limit for $\Phi_{\rm HO}$ while the conformal invariance of the transverse momentum ${\bf k}_{\perp}$ 
doesn't hold in the chiral limit for $\Phi_{\rm PL}$. This distinguishes the behavior of the chiral limit between $\Phi_{\rm HO}$ and $\Phi_{\rm PL}$
and leads to the difference in the chiral limit for $\phi^{P}_{3;M}(x)$ depending on which LF model wave function is applied for the computation.
We present more details of the chiral limit behaviors for each case of the LF trial wave functions discussed in this work.

In our previous analyses of twist-2 and pseudoscalar twist-3 DAs of a pseudoscalar meson~\cite{CJ2015}
and  the chirality-even twist-2 and twist-3 DAs of a vector meson~\cite{CJ_V14},
we have shown that the results in the standard LFQM is obtained by 
the mapping 
of the LF vertex function $\chi$ in BS model 
into
our LFQM wave function
$\Phi_R$ as follows~(see Eq.~(35) in~\cite{CJ2015} or Eq.~(49) in~\cite{CJ_V14})
\be\label{Deq:12}
 \sqrt{2N_c} \frac{ \chi(x,{\bf k}_\perp) } {1-x}
 \to \frac{\Phi_R (x,{\bf k}_\perp) }
 {\sqrt{{\bf k}^2_\perp + {\mathfrak A}^2}}, \; m_M \to M_0,
 \ee
where ${\mathfrak A}=(1-x)m_q + x m_{\bar q}$ 
and $m_M\to M_0$ implies that the physical mass $m_M$ included in the integrand of BS
amplitude has to be replaced with the invariant mass $M_0$ since the results in the standard LFQM
are obtained from the requirement of all constituents being on their respective mass shell.
The correspondence in Eq.~(\ref{Deq:12}) is valid again in this analysis of a
pseudotensor twist-3 DA $\phi^{\sigma}_{3;M}(x)$.

We now apply the same mapping to both $\phi^{P}_{3;M}(x)$ in Eq.~(\ref{Deq:5.1})  and
$\phi^{\sigma}_{3;M}(x)$ in Eq.~(\ref{Deq:5.2})
to obtain them in our LFQM as follows:
\bea\label{Deq:13}
 \phi^{P}_{3;M}(x)
 &=& \frac{\sqrt{2N_c}}{f_M\mu_M}
 \int \frac{d^2{\bf k}_\perp}{16\pi^3}
 \frac{\Phi_R(x,{\bf k}_\perp)}{\sqrt{{\bf k}^2_\perp + {\mathfrak A}^2}}
 [M^2_0 -(m_q - m_{\bar q})^2],
\eea
and
\bea\label{Deq:14}
 \phi^{\sigma}_{3;M}(x)
 &=& \frac{6\sqrt{2N_c}}{f_M\mu_M}
 \int \frac{d^2{\bf k}_\perp}{16\pi^3} \int^x_0 dx'
 \frac{\Phi_R(x',{\bf k}_\perp)}{\sqrt{{\bf k}^2_\perp + {\mathfrak A'}^2}}
 [(1-2x') M'^2_0 + m^2_q - m^2_{\bar q}],
\eea
respectively, where ${\mathfrak A'}={\mathfrak A}(x \to x')$.
It is remarkable to observe that both the zero-mode contribution and the instantaneous contribution are absorbed into the LF on-mass-shell constituent quark and antiquark
contribution as shown in Eqs.(\ref{Deq:13}) and (\ref{Deq:14}).

For the point of view of QCD, one should note that the quark-antiquark DAs of a hadron depend on the scale $\mu$ that may separate nonperturbative
and perturbative regimes. In our LFQM, we can associate $\mu$ with the transverse integration cutoff via $|{\bf k}_\perp|\leq \mu$. The
dependence on the scale $\mu$ is then consistently given by the QCD evolution equation~\cite{BL80}, while the DAs at a certain low scale
can be obtained by the necessary nonperturbative input from LFQM. As the cutoff dependence becomes marginal beyond a certain nonperturbative cutoff scale, 
the Gaussian (or HO) and PL wave functions  given by Eqs.~(\ref{QM2}) and~(\ref{QM2_PL}) are allowed to perform
the integral up to infinity without any appreciable loss of accuracy. 

\section{Numerical Results}
\label{sec:III}
\begin{table}[t]
\caption{
Model parameters  for the Gaussian wave function with the linear and HO confining potentials~\cite{CJ_99,CJ_DA,Choi07}
and for the power-law wave function~\cite{schlumpf}. $q=u$ and $d$.}
\label{t1}
\begin{tabular}{ccccc} \hline\hline
Model &$m_q$ (GeV) & $m_s$ (GeV) & $\beta_{q{\bar q}}$ (GeV) & $\beta_{q{\bar s}}$  (GeV)\\
\hline
Linear &0.22 & 0.45 & 0.3659 & 0.3886  \\
HO &  0.25 & 0.48 & 0.3194 & 0.3419  \\
Power-Law &0.25 & 0.37 & 0.335 & 0.41  \\
\hline\hline
\end{tabular}
\end{table}

In the numerical computations, we use the linear and HO~confining potential model parameters
for the Gaussian wave function
given in Table I, which were obtained from the calculation of meson mass spectra using the variational principle
in our LFQM~\cite{CJ_99,CJ_DA,Choi07}.
For the sensitivity analysis depending on the form of the model wave functions, we also use the PL wave function with the model parameters adopted from Ref.~\cite{schlumpf}.
Since our numerical results for the twist-2 $\phi^A_{2;M}(x)$ and twist-3  $\phi^P_{3;M}(x)$  of $\pi$ an $K$ mesons were presented
in our previous works~\cite{CJ_DA,CJ2015},
we shall focus on the calculation of the twist-3 $\phi^\sigma_{3;M}(x)$ of $\pi$ 
and
$K$ mesons together with
some new results for $\phi^P_{3;M}(x)$ including the PL wave function  in this work.

Defining the LF wave function $\psi^{P(\sigma)}_{3; M}(x,{\bf k}_\perp)$
for the twist-3 pseudoscalar (pseudotensor) channel as
\be\label{N1}
\phi^{P(\sigma)}_{3;M}(x) = \int^\infty_0 d^2{\bf k}_\perp \psi^{P(\sigma)}_{3;M}(x,{\bf k}_\perp),
\ee
the $n$-th transverse moment is obtained by
\be\label{N2}
\la {\bf k}^n_\perp \ra^{P(\sigma)}_M = \int^\infty_0 d^2{\bf k}_\perp
\int^1_0 dx \psi^{P(\sigma)}_{3;M}(x,{\bf k}_\perp) {\bf k}^n_\perp.
\ee

For the pion case, our results of the second transverse moments for 
$\psi^{P}_{3; \pi}(x,{\bf k}_\perp)$ and $\psi^{\sigma}_{3; \pi}(x,{\bf k}_\perp)$
obtained from the linear~[HO] parameters
are $\la {\bf k}^2_\perp \ra^P_\pi = (553~{\rm MeV})^2 [(480~{\rm MeV})^2]$ and
$\la {\bf k}^2_\perp \ra^\sigma_\pi = (481~{\rm MeV})^2 [(394~{\rm MeV})^2]$, respectively.
For the kaon case, we obtain
$\la {\bf k}^2_\perp \ra^P_K = (582~{\rm MeV})^2 [(510~{\rm MeV})^2] $ and
$\la {\bf k}^2_\perp \ra^\sigma_K = (481~{\rm MeV})^2 [(428~{\rm MeV})^2] $
for the linear~[HO] parameters, respectively.
Since the PL wave function given by Eq.~(\ref{QM2_PL}) is not enough power suppressed to give finite transverse moments
unless the transverse integration
cutoff is performed, we do not estimate them for the PL wave function case.

\begin{figure}
\begin{center}
\includegraphics[height=7cm, width=7cm]{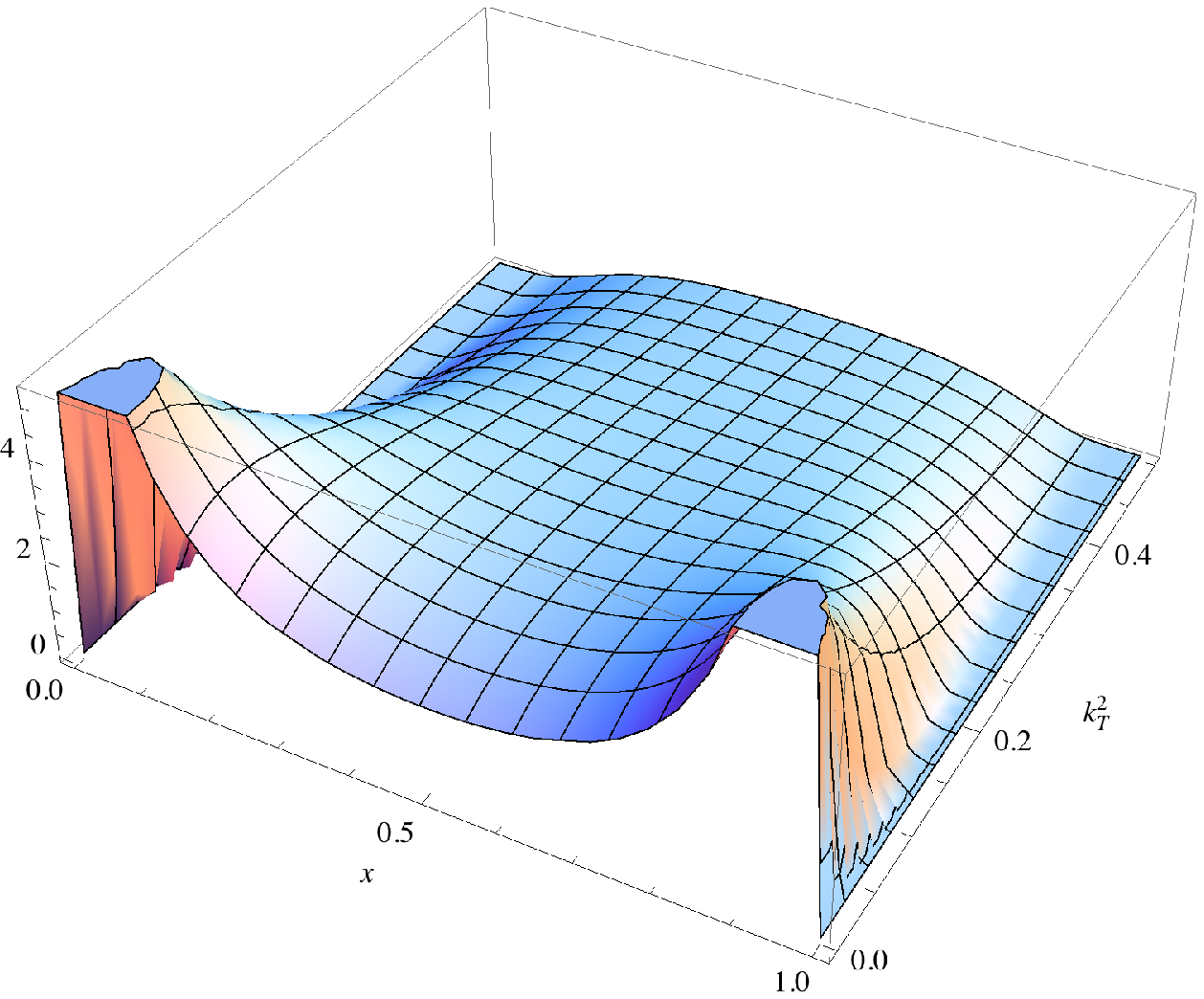}
\hspace{0.5cm}
\includegraphics[height=7cm, width=7cm]{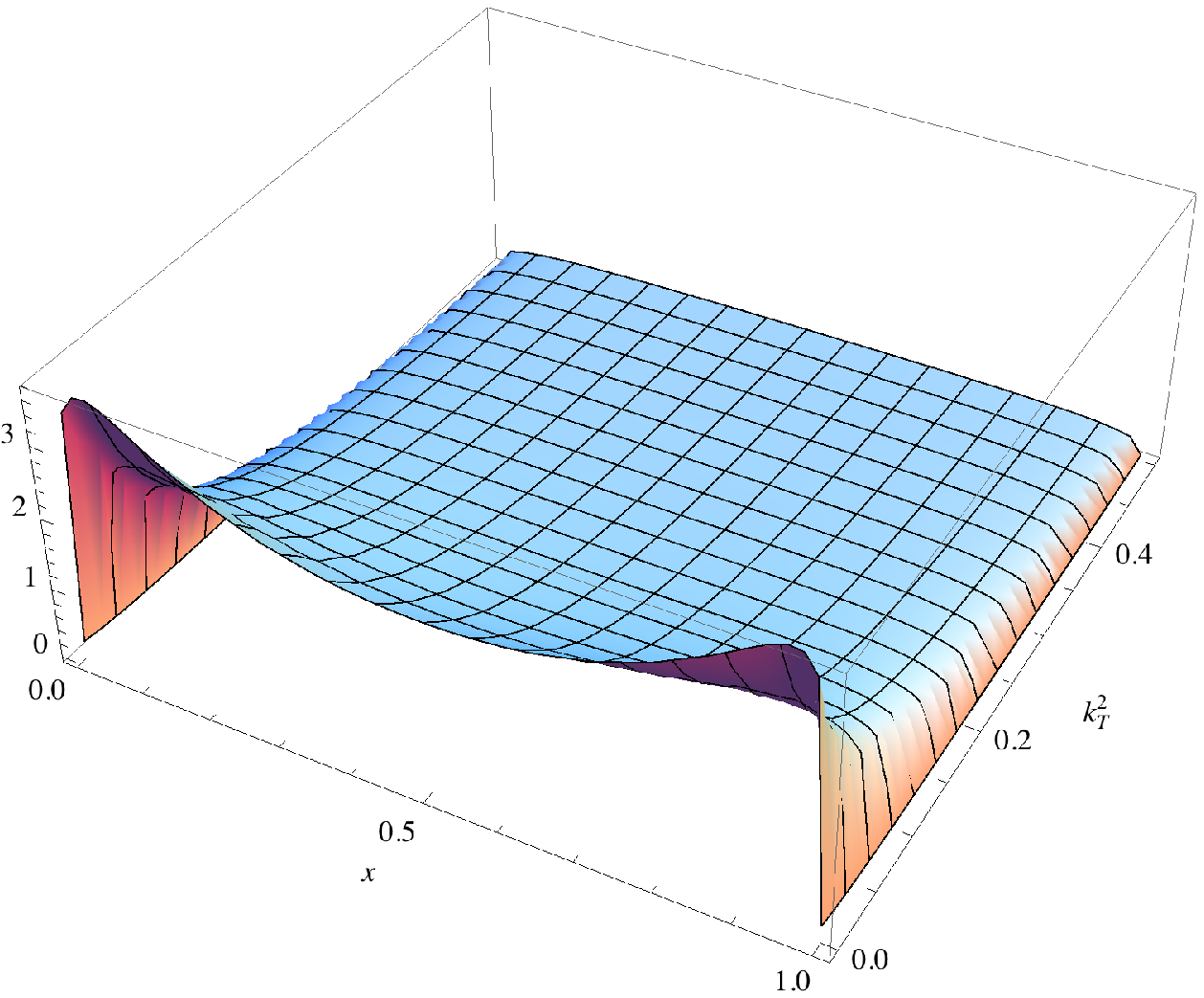}
\hspace{0.5cm}
\includegraphics[height=7cm, width=7cm]{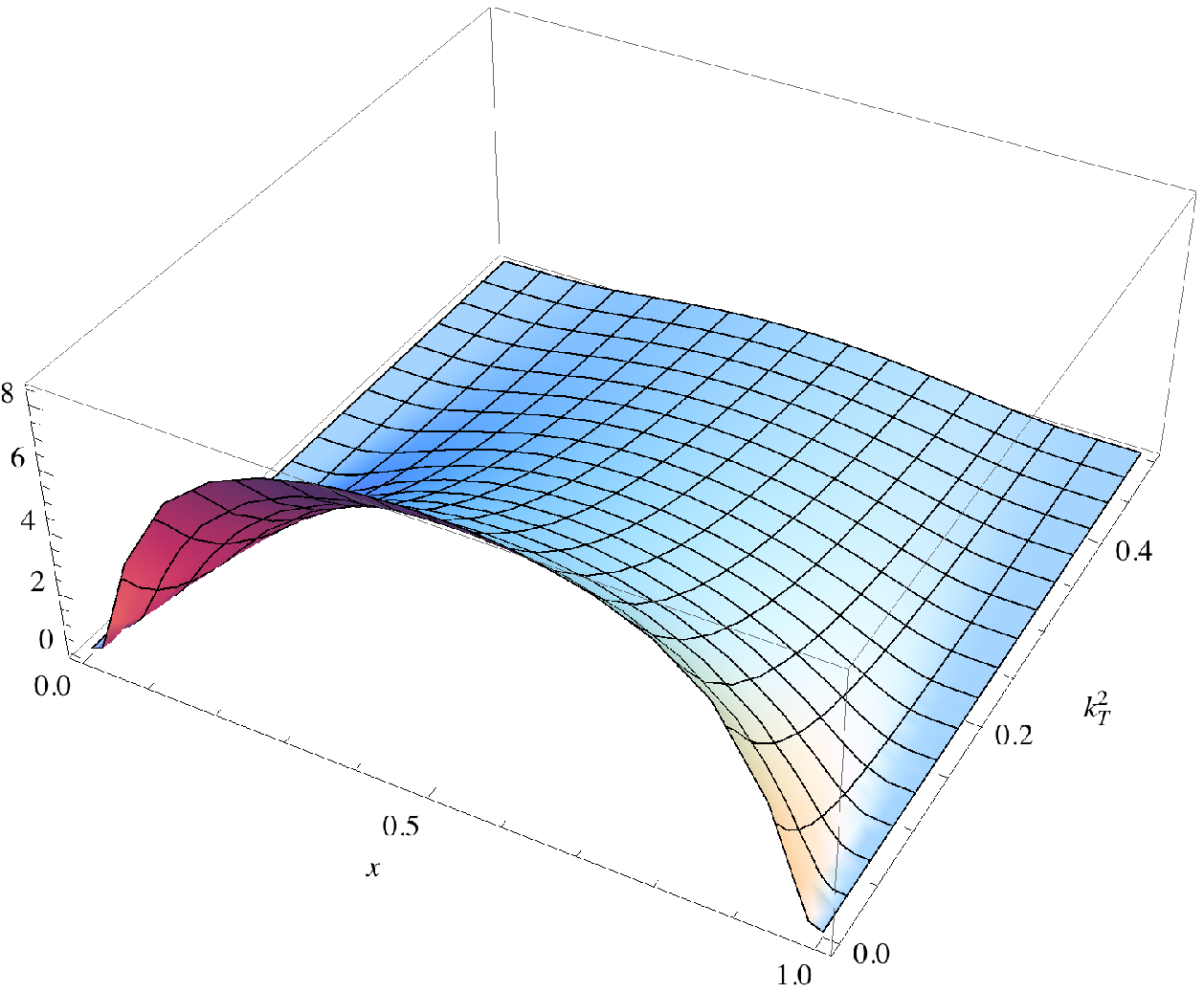}
\hspace{0.5cm}
\includegraphics[height=7cm, width=7cm]{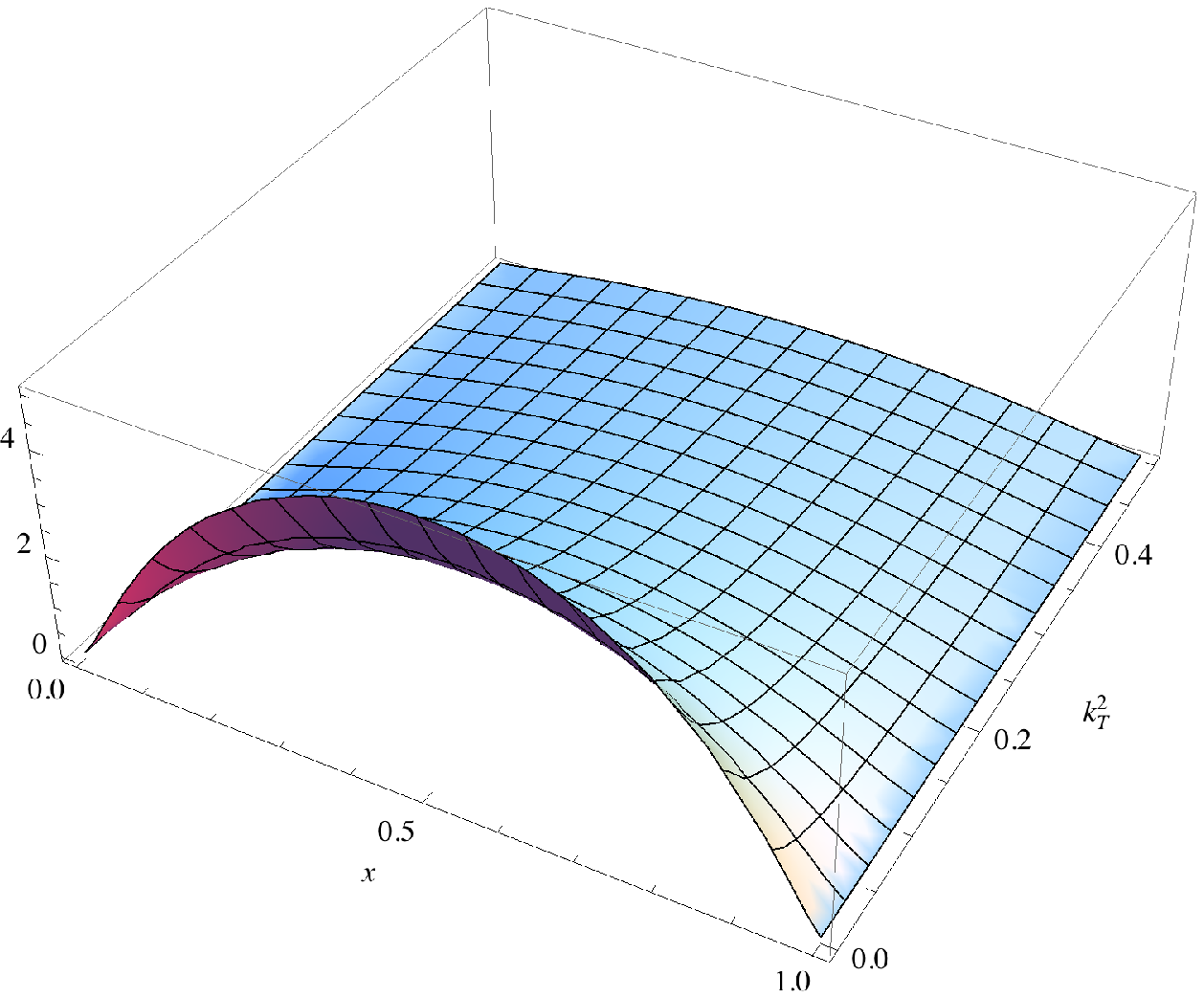}
\caption{\label{fig2} 3D plots for $\psi^P_{3;\pi}(x,{\bf k}_\perp)$ (upper panel) and $\psi^\sigma_{3;\pi}(x,{\bf k}_\perp)$ (lower panel) obtained
from the HO (left panel) and the PL (right panel) wave functions, respectively.}
\end{center}
\end{figure}
%

\begin{figure}
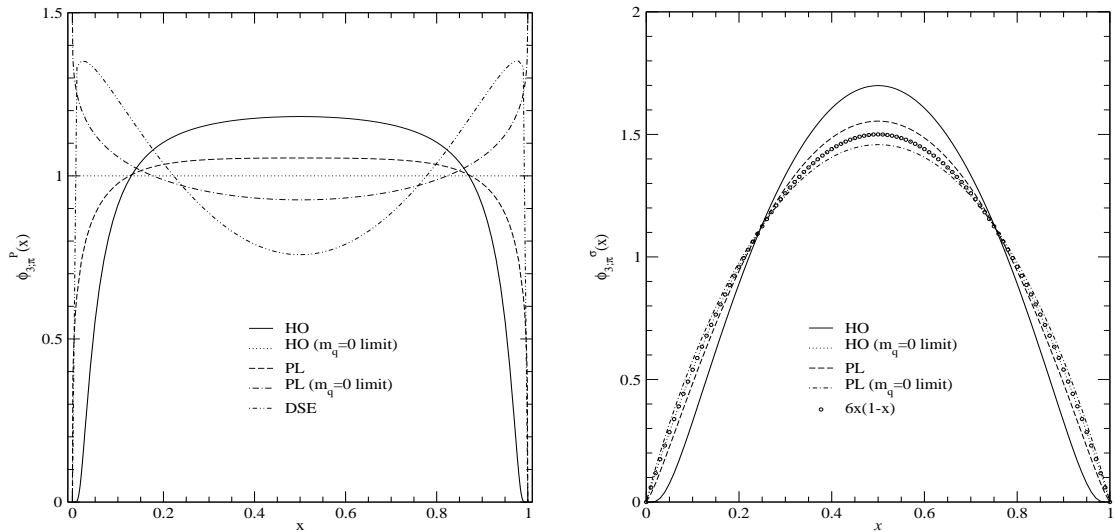

\vspace{0.5cm}
\begin{center}
\includegraphics[height=7cm, width=7cm]{Fig3a.eps}
\hspace{0.5cm}
\includegraphics[height=7cm, width=7cm]{Fig3b.eps}
\caption{\label{fig3} The twist-3 DAs $\phi^P_{3;\pi}(x)$  (left panel) and $\phi^\sigma_{3;\pi}(x)$ (right panel) of pion.}
\end{center}
\end{figure}

Fig.~\ref{fig2} shows the 3D plots for the twist-3 pion LF wave functions 
$\psi^{P}_{3; \pi}(x,{\bf k}_\perp)$ (upper panel) and
$\psi^{\sigma}_{3; \pi}(x,{\bf k}_\perp)$ (lower panel) 
obtained from the Gaussian wave functions with HO model parameters (left panel) and the PL
wave functions (right panel), respectively. 
For the case of pseudoscalar $\psi^{P}_{3; \pi}(x,{\bf k}_\perp)$, it shows the concave
shape for low ${\bf k}^2_\perp$ for both Gaussian and PL wave functions but 
its DA $\phi^{P}_{3; \pi}(x)$ after the ${\bf k}_\perp$-integration up to infinity
shows rather convex shape in the central region of $x$ as we show in Fig.~\ref{fig3}.
On the other hand, for the case of pseudotensor $\psi^{\sigma}_{3; \pi}(x,{\bf k}_\perp)$,
it shows the convex shape for any value of ${\bf k}^2_\perp$ regardless the choice of the wave functions. For both pseudoscalar and
pseudotensor channels, the PL wave functions have more high momentum tails than the corresponding Gaussian wave functions for
$|{\bf k}_\perp| \geq 1$ GeV. Thus, the PL wave functions are rather sensitive to the transverse momentum cutoff values. 
We also should note that $\psi^{P}_{3; \pi}(x,{\bf k}_\perp)$ is much more sensitive to the choice of the LF wave functions than 
$\psi^{\sigma}_{3; \pi}(x,{\bf k}_\perp)$. This may lead to different asymptotic behaviors for different LF wave functions in the chiral symmetry limit.

We show in Fig.~\ref{fig3} the corresponding two-particle twist-3 pion $\phi^P_{3;\pi}(x)$  (left panel) and $\phi^\sigma_{3;\pi}(x)$ (right panel)
obtained from  the nonzero constituent quark masses using Gaussian wave functions with HO (solid lines) model parameters
and PL wave functions (dashed lines). We also plot our results in the chiral symmetry $(m_{u(d)}\to 0)$ limit for both Gaussian (dotted lines)
and PL (dot-dashed lines) wave functions and compare them with the chiral-limit prediction of DSE approach
employing the dynamical chiral symmetry breaking (DCSB) improved (BD) kernels~\cite{DSE2015} (double-dot-dashed line) as well as the asymptotic result $6x(1-x)$ 
for the case $\phi^\sigma_{3;\pi}(x)$.
Our results  for both $\phi^P_{3;\pi}(x)$  and $\phi^\sigma_{3;\pi}(x)$  are normalized without the momentum cutoff (i.e. $|{\bf k}_\perp|\to\infty$). 

For the $\phi^P_{3;\pi}(x)$ case in Fig.~\ref{fig3}, our results with nonzero constituent quark masses show rather convex shapes for both Gaussian and PL
wave functions but they show quite different end point behaviors, i.e. the end points are more enhanced for the PL wave function than
the Gaussian wave function. The difference between the two wave functions are more drastic in the chiral symmetry limit, where the result of Gaussian
wave function  reproduces the  result $\phi^P_{3;\pi}(x)\to 1$  anticipated from the QCD's conformal limit~\cite{BF}
but the result of PL wave function shows the concave shape similar to the result of DSE approach~\cite{DSE2015}, in which
the following asymptotic form was obtained: $\phi^P_{3;\pi}(x)\to 1 + (1/2) C^{(1/2)}_2(2x-1)$.
This rebuts the remark made in Ref.\cite{DSE2015} that our LFQM has curvature of the opposite sign
on almost the entire domain of support in conflict with a model-independent prediction of QCD.
We have shown in our previous works\cite{CJ_V14,CJ2015} that our LFQM is indeed consistent with the nature of chiral symmetry in QCD.
While the authors in~\cite{DSE2015} explained that the difference, i.e. $(1/2)C^{(1/2)}_2(2x-1)$ term in chiral symmetry limit, may come from the
mixing effect between the two- and three-particle twist-3 amplitudes, we observe the similar difference taking the power-law type LF wave function 
in which the transverse momentum dependence cannot be factorized from the scale independent parameter dependence.  
Especially, we find that the end point behaviors of $\phi^P_{3;\pi}(x)$ also affect the asymptotic form in the chiral symmetry limit.
The cutoff dependent behaviors of $\phi^P_{3;\pi}(x)$ obtained from both Gaussian and PL wave functions are also presented in Ref.~\cite{FB2015}, where
the concave shape for the Gaussian wave function can also be seen with the cutoff scale $\mu=1$ GeV or less being taken but the
cutoff dependence was shown to be more sensitive for the PL wave function than the Gaussian one.

For the $\phi^\sigma_{3;\pi}(x)$ case in Fig.~\ref{fig3},  our results with nonzero constituent quark masses for both Gaussian (solid line) and PL (dashed line)
show again different end point behaviors, i.e. the end points are more enhanced for the PL wave function than
the Gaussian wave function.
However, in the chiral symmetry limit,  Gaussian (dotted line) and PL (dot-dashed line) wave functions show very
similar shapes each other. Thus, the degree of conformal symmetry breaking depends on the channel of DAs, $\phi^P_{3;\pi}(x)$ vs. $\phi^\sigma_{3;\pi}(x)$.
As expected, the result from Gaussian wave function reproduces exactly the asymptotic form $6x(1-x)$.
The same chiral-limit behavior was also obtained from the DSE approach~\cite{DSE2015}.
As one can see from Fig.~\ref{fig3}, the twist-3 pseudoscalar $\phi^P_{3;\pi}(x)$  
is more sensitive to the shape of
the model wave functions (Gaussian vs. PL) than the twist-3 pseudotensor $\phi^\sigma_{3;\pi}(x)$.
It is quite interesting to note in the chiral symmetry limit that while $\phi^P_{3;\pi}(x)$ is sensitive to the shapes of model wave functions,
$\phi^\sigma_{3;\pi}(x)$ is insensitive to them.

\begin{figure}
\begin{center}
\includegraphics[height=7cm, width=7cm]{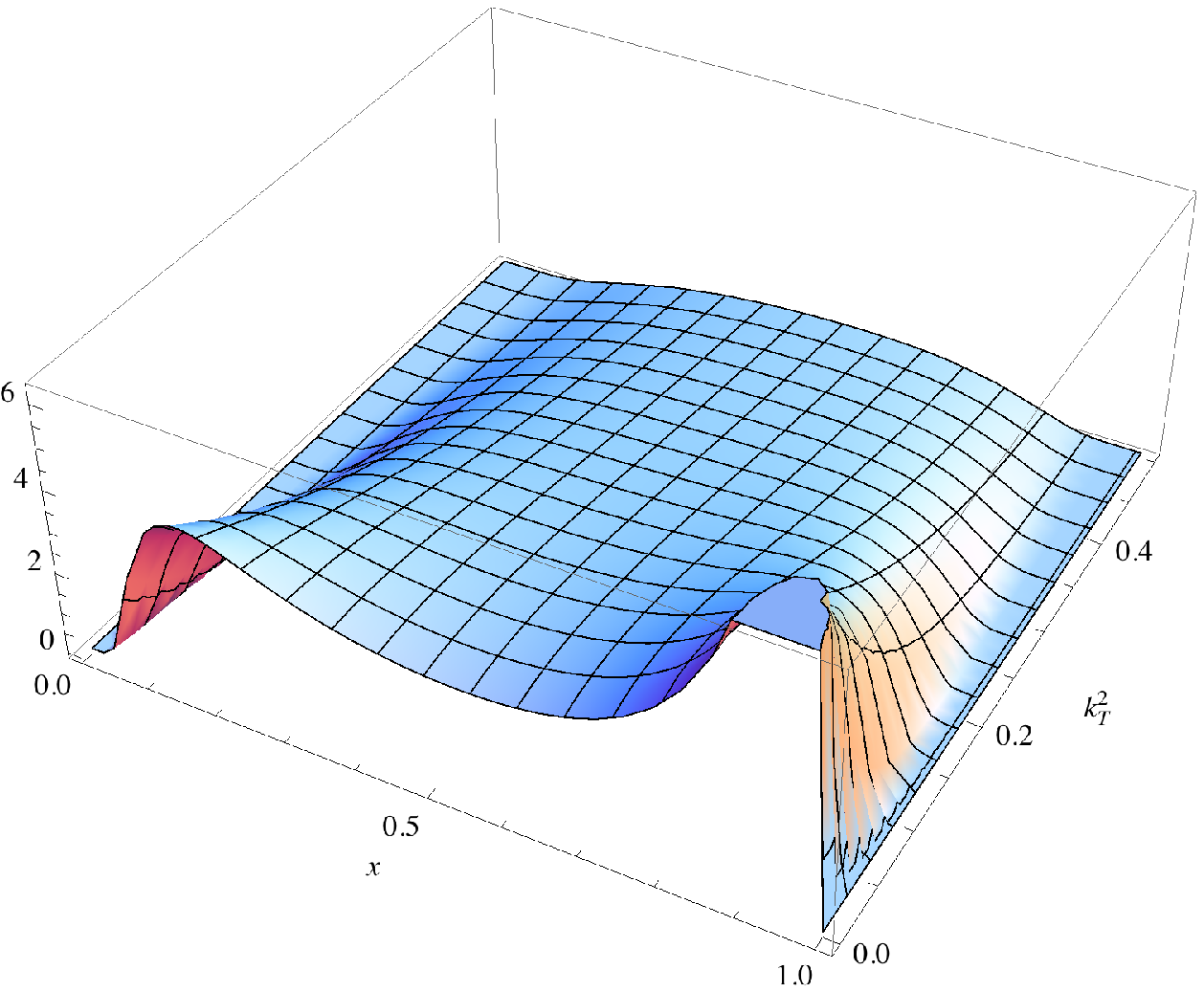}
\hspace{0.5cm}
\includegraphics[height=7cm, width=7cm]{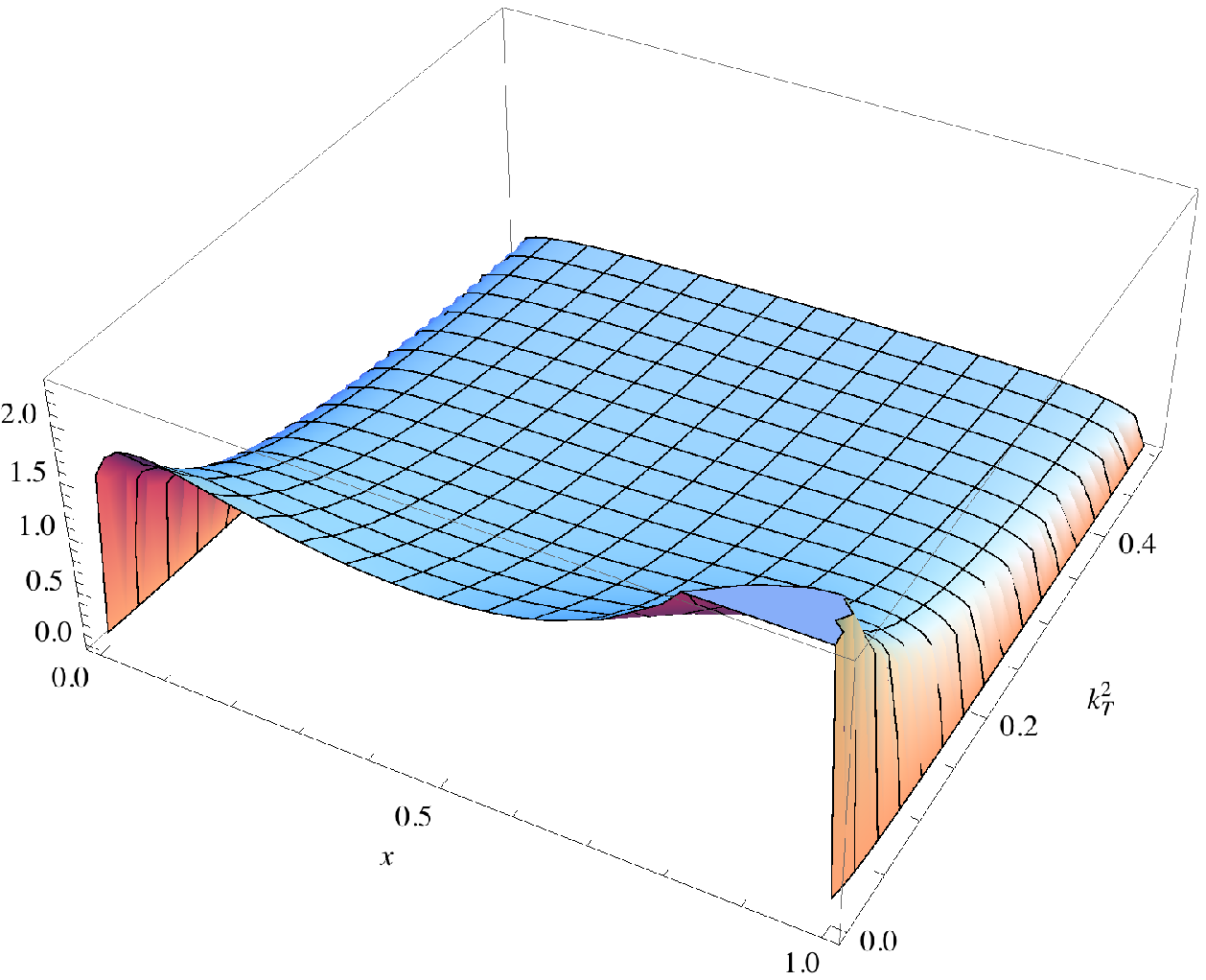}
\hspace{0.5cm}
\includegraphics[height=7cm, width=7cm]{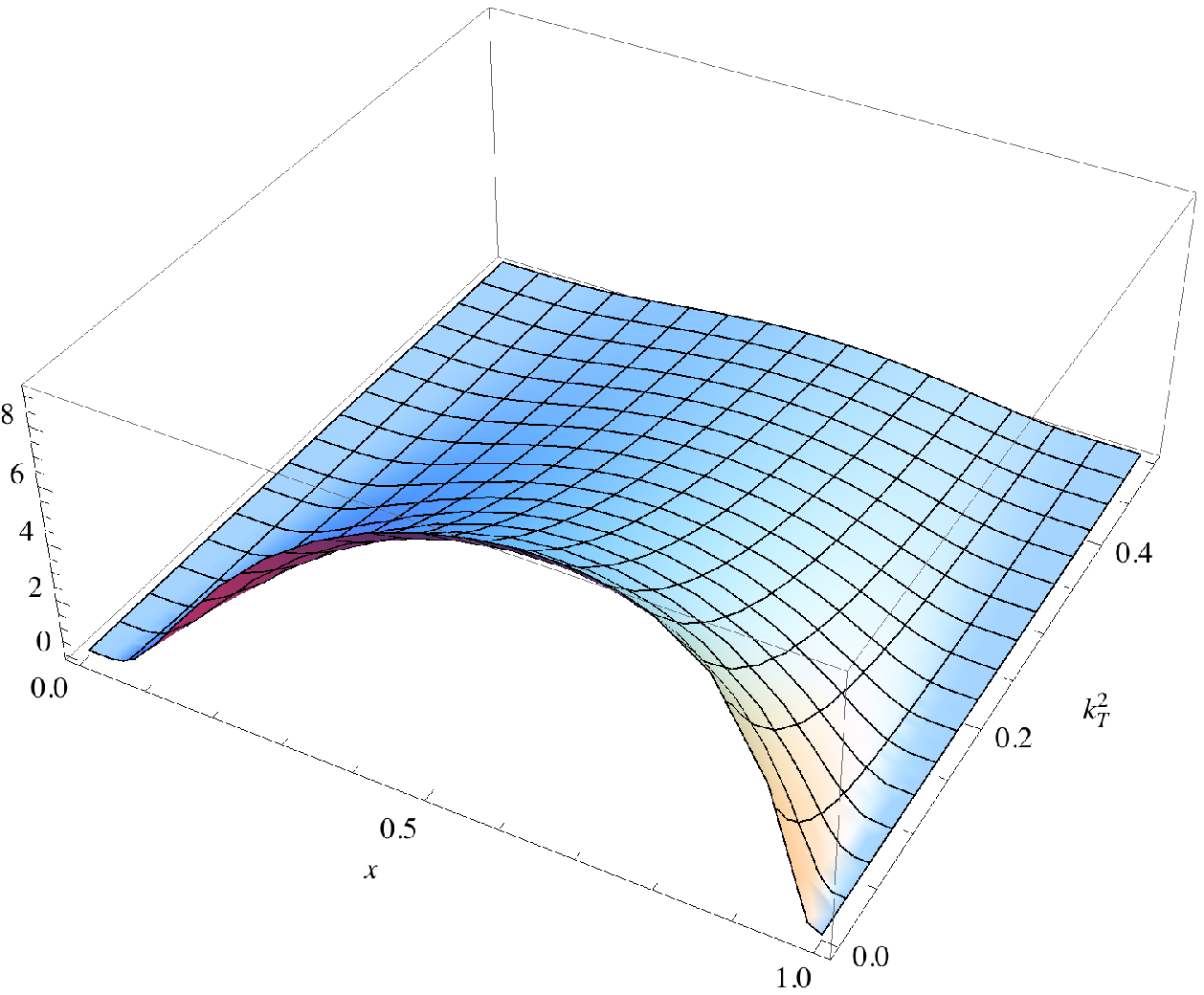}
\hspace{0.5cm}
\includegraphics[height=7cm, width=7cm]{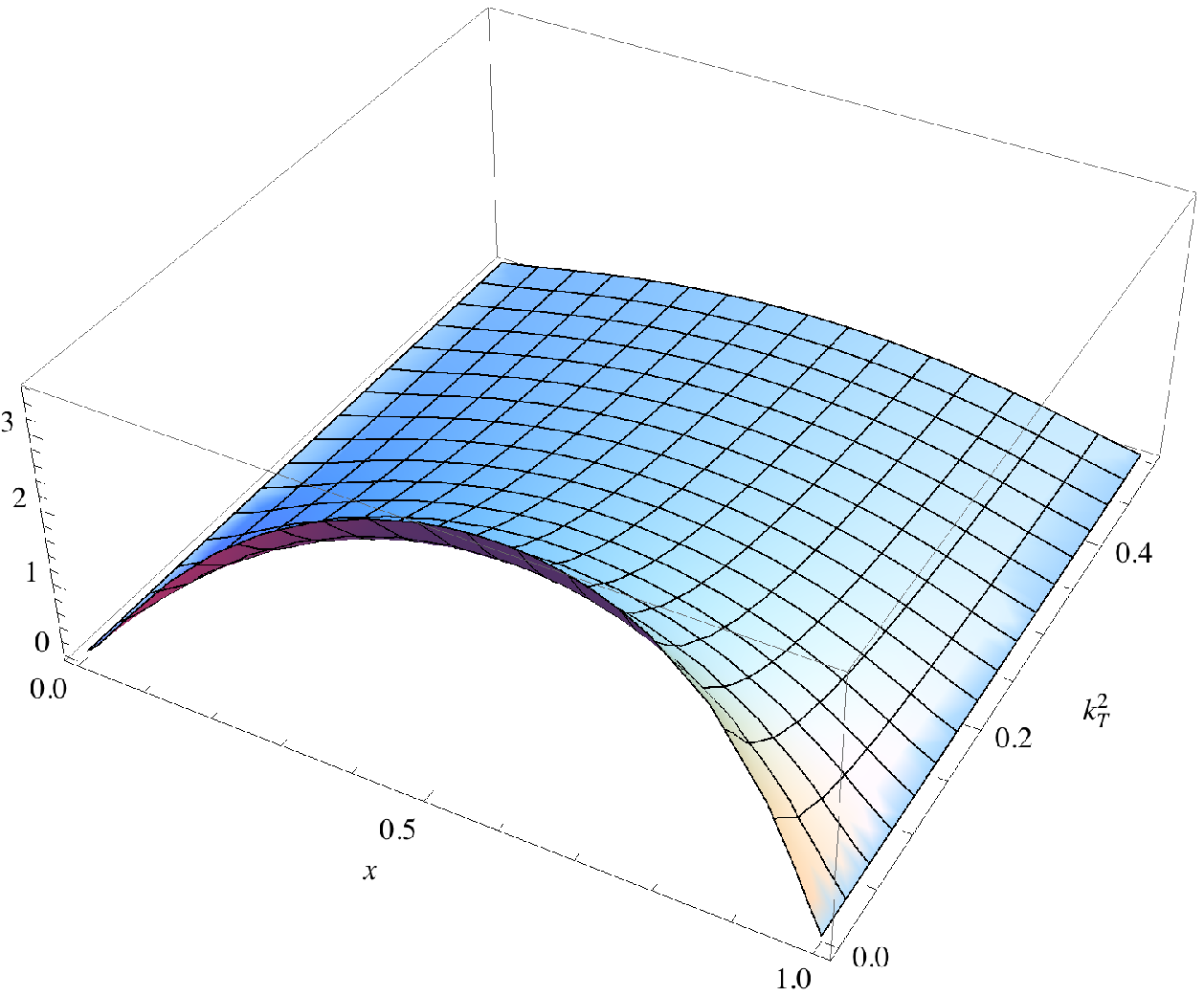}
\caption{\label{fig4} The 3D plots for $\psi^P_{3;K}(x,{\bf k}_\perp)$ (upper panel)
and $\psi^\sigma_{3;K}(x,{\bf k}_\perp)$ (lower panel) obtained
from the HO (left panel) and the PL (right panel) wave functions.}
\end{center}
\end{figure}
%

Fig.~\ref{fig4} shows the 3D plots for the twist-3 kaon LF wave functions 
$\psi^{P}_{3; K}(x,{\bf k}_\perp)$ (upper panel) and
$\psi^{\sigma}_{3; K}(x,{\bf k}_\perp)$ (lower panel) obtained from the Gaussian wave functions with HO model parameters (left panel) and the PL
wave functions (right panel), respectively.  For the kaon case, we assign the momentum fractions $x$ for $s$-quark
and $(1-x)$ for the light $u(d)$-quark.
Due to the SU(3) flavor-symmetry breaking effect, the twist-3 kaon LF wave functions are distorted in favor of the heavier $s$-quark. Other than the SU(3) flavor-symmetry breaking effect, the general behavior is similar to the pion case.

\begin{figure}
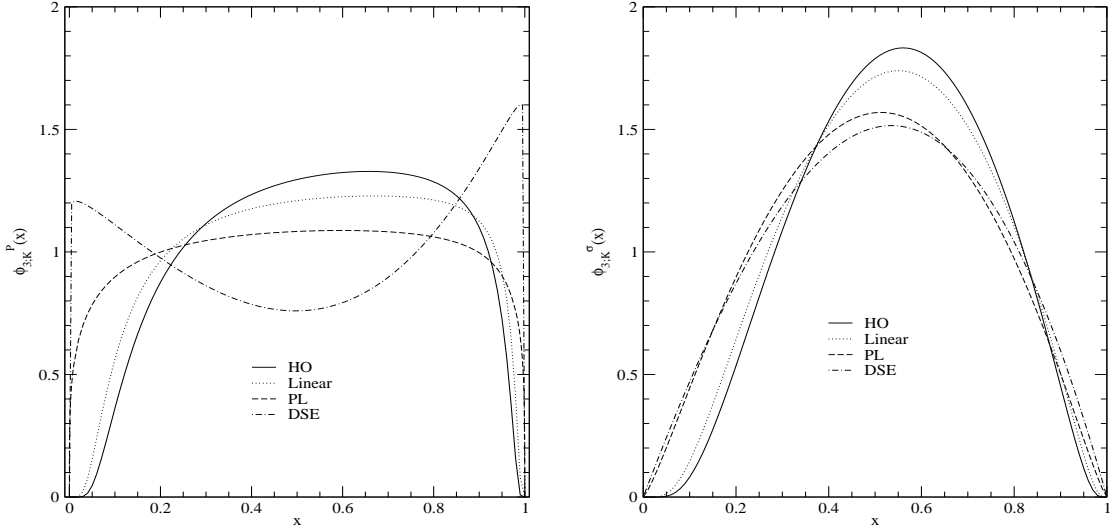

\vspace{0.5cm}
\begin{center}
\includegraphics[height=7cm, width=7cm]{Fig5a.eps}
\hspace{0.5cm}
\includegraphics[height=7cm, width=7cm]{Fig5b.eps}
\caption{\label{fig5} The twist-3 DAs $\phi^P_{3;K}(x)$  (left panel) and $\phi^\sigma_{3;K}(x)$ (right panel) of pion.}
\end{center}
\end{figure}

We show in Fig.~\ref{fig5} the corresponding two-particle twist-3 kaon $\phi^P_{3;K}(x)$  (left panel) and $\phi^\sigma_{3;K}(x)$ (right panel)
obtained from Gaussian wave functions with HO (solid lines) and Linear(dotted lines) model parameters and PL wave functions (dashed lines).
We also  compare our results with the prediction of 
DSE approach
employing the dynamical chiral symmetry breaking-improved (BD) kernels~\cite{DSE2015} (dot-dashed line).
Our results  for both $\phi^P_{3;K}(x)$  and $\phi^\sigma_{3;K}(x)$  are normalized without the transverse momentum cutoff.
In both pseudoscalar and pseudotensor twist-3 kaon DAs, the difference between the HO and Linear model parameters
using the same Gaussian wave functions is less significant than the difference between the Gaussian and PL  wave functions.
On the other hand, the SU(3) flavor-symmetry breaking effect is more pronounced in the Gaussian wave function than the PL wave function.
As in the case of pion, while some disagreements between our LFQM prediction and DSE prediction are seen in $\phi^P_{3;K}(x)$,
some agreements between them can also be seen in $\phi^\sigma_{3;K}(x)$. Especially, for the pseudotensor DA $\phi^\sigma_{3;K}(x)$,
our prediction from PL wave function is in good agreement to the result from DSE approach including the end points behaviors.
 As was discussed in~\cite{DSE2015}, the SU(3) flavor-symmetry breaking effect of two-particle twist-3 kaon DAs may be quantified by considering a ratio, viz.
\be\label{SU3B}
\delta_{\phi^{P(\sigma)}_{3;K}}=\frac{\int^{1/2}_{0} d{\bar{x}}\phi^{P(\sigma)}_{3;K}( 1-\bar{x})}
{\int^{1/2}_{0} dx\phi^{P(\sigma)}_{3;K}(x)},
\ee
where ${\bar x}=1-x$. We obtain
$\delta_{\phi^{P}_{3;K}}=(1.28, 1.38, 1.06)$
and $\delta_{\phi^{\sigma}_{3;K}}=(1.33, 1.43, 1.05)$ for (Linear, HO, PL) parameters, respectively.
The same formula as in Eq.~(\ref{SU3B}) should hold for twist-2 DA ($\phi^P_{2;K}$)~\cite{CJ_DA,CJ2015},
and we obtain
the ratio as  $\delta_{\phi^{P}_{2;K}}=(1.15, 1.28, 1.16)$ for (Linear, HO, PL) parameters.
Our results should be compared with the DSE approach~\cite{DSE2015} results
using two different procedures, i.e. rainbow-ladder (RL) truncation and the DCSB-improved (DB) kernels:
$\delta_{\phi^{P}_{3;K}}=\delta_{\phi^{\sigma}_{3;K}}=(1.28, 1.12)$ for (RL, DB) and  $\delta_{\phi^{P}_{2;K}}=1.14$ for DB, respectively.
As one can see from our results,
the SU(3) flavor-symmetry breaking effect is larger for Gaussian wave function than for PL wave function. Overall
our results from the PL wave
function agree quantitatively with the DSE results from DCSB-improved kernels.
Regarding on the flavor symmetry breaking effect, 
our LFQM results~\cite{CJ_DA,Choi07} of leptonic decay constant ratios $f_K/f_\pi =1.24[1.18]$ and $f_{B_s}/f_B=1.24 [1.32]$ obtained 
from Gaussian wave functions with Linear [HO] parameters
can also be compared with the experimental data
$f_K/f_\pi =1.22$~\cite{PDG14} and the recent unquenched lattice-QCD  $f_{B_s}/f_B=1.22(8)$~\cite{LQCD15},
respectively.

The twist-3 pseudoscalar DA $\phi^P_{3;M}(x)$ and pseudotensor DA $\phi^\sigma_{3;M}(x)$ are usually expanded in terms of the
Gegenbauer polynomials $C^{1/2}_n$ and $C^{3/2}_n$, respectively, as follows~\cite{NK06}:
\bea\label{Ge1}
\phi^P_{3;M} &=&  \sum^{\infty}_{n=0} a^{P}_{n,M}C^{1/2}_n(2x-1),
 \nonumber\\
 \phi^\sigma_{3;M} &=& 6 x(1-x) \sum^{\infty}_{n=0} a^{\sigma}_{n,M}C^{3/2}_n(2x-1).
\eea
The coefficients
$a^{P(\sigma)}_{n,M}$ are called the Gegenbauer moments and can be obtained by
\bea\label{Gel2}
a^P_{n,M}(x) &=& (2n+1) \int^1_0 dx C^{1/2}_n(2x-1)\phi^P_{3;M}(x),
\nonumber\\
a^\sigma_{n,M}(x) &=& \frac{4n+6}{3n^2+9n+6}\int^1_0 dx C^{3/2}_n(2x-1)\phi^\sigma_{3;M}(x),
\eea
using the orthogonal
condition for the Gegenbauer polynomials
\be\label{Ge2}
\int^1_0 dx [x(1-x)]^{l-1/2} C^l_m(2x -1) C^l_n (2x -1)
=\frac{\pi 2^{1-4l}\Gamma(2l+n)}{n!(n+l)\Gamma^2(l)}\delta_{mn}.
\ee
The Gegenbauer moments with $n>0$ describe how much the DAs deviate from the asymptotic one.
In addition to the Gegenbauer moments, one can also define the expectation value of the longitudinal
momentum, so-called $\xi$-moments, as follows:
\be\label{Ge3}
\la\xi^n\ra^{P(\sigma)}_M = \int^1_0 dx \xi^n \phi^{P(\sigma)}_{3;M}(x),
\ee
where $\xi=2x-1$.

\begin{table}[t]
\caption{The Gegenbauer moments and $\xi$ moments of  twist-3 pion DAs obtained from the
linear and HO potential models compared with other model estimates. }
\label{t2}
\renewcommand{\tabcolsep}{1pc} 
\begin{tabular}{@{}lcccccc} \hline\hline
Models &  $a^{\sigma}_{2,\pi}$ & $a^{\sigma}_{4,\pi}$ & $a^{\sigma}_{6,\pi}$
& $\la\xi^2\ra^{\sigma}_\pi$ & $\la\xi^4\ra^{\sigma}_\pi$ & $\la\xi^6\ra^{\sigma}_\pi$ \\
\hline
HO & -0.1155 & -0.0268 & -0.0046 & 0.1604 & 0.0565 & 0.0263 \\
Linear & -0.0803 &  -0.0256 &  -0.0082 & 0.1725 &  0.0647 & 0.0318 \\
PL       & -0.0375 & -0.0092 & -0.0031 & 0.1871 & 0.0762 & 0.0406 \\
SR~\cite{BBL} & 0.0979&  -0.0016 &  -0.0011 & 0.2325 &  0.1075 & 0.0624 \\
DSE~\cite{DSE2015} & $\cdots$ &  $\cdots$ &  $\cdots$ & 0.20 & 0.085 & 0.047 \\
$\chi$QM~\cite{NK06} & -0.0984 &  -0.0192 &  -0.0037 & 0.1663 &  0.0612 & -0.0015 \\
$6x(1-x)$ & $\cdots$ &  $\cdots$ &  $\cdots$ & 0.20 & 0.086 & 0.048 \\
\hline\hline
\end{tabular}
\end{table}

Since we calculated Gegenbauer- and $\xi$- moments of the pseudoscalar 
twist-3 $\phi^P_{3;M}(x)$ as well as the twist-2 $\phi^A_{2;M}(x)$
in our previous works~\cite{CJ_DA,CJ2015}, we do not
list them here.

In Table~\ref{t2}, we list the calculated Gegenbauer- and $\xi$- moments of 
 the pseudotensor twist-3 pion $\phi^\sigma_{3;\pi}(x)$
obtained from the Gaussian wave function
with linear and HO potential models and PL wave function.
We also compare our results  with other model predictions, e.g.
QCD sum rules (SR)~\cite{BBL}, DSE approach~\cite{DSE2015} and the chiral quark model ($\chi$QM)~\cite{NK06}.
As expected from the isospin symmetry, all odd Gegenbauer and $\xi$ moments are zero.
It is interesting to note that the sign of
$a^{\sigma}_{2,\pi}$ is negative from our LFQM and $\chi$QM predictions but is positive for QCDSR prediction.
Larger positive value of $a^{\sigma}_{2,\pi}$ leads to more flat shape of DA but the larger negative value leads to more narrower
shape of DA as one can see from Fig.~\ref{fig3}.
Knowing our LFQM results from the HO model are exact to the asymptotic result in the chiral-symmetry limit as shown in
Fig.~\ref{fig3}, i.e. $[\la\xi^2\ra^{\sigma}_\pi]_{\rm HO}=[\la\xi^2\ra^{\sigma}_\pi]_{\rm asy}$ in $m_q\to 0$ limit, one can see
that the $\xi$-moments are reduced when the chiral symmetry is broken. We also should note
for the same reason
that our LFQM results are in good agreement with DSE results in the chiral-symmetry limit of $\phi^\sigma_{3;\pi}(x)$.

\begin{table}[t]
\caption{The Gegenbauer moments and $\xi$ moments of twist-2 and twist-3 $K$ meson DAs obtained from the
linear and HO potential models compared with other model estimates. }
\label{t3}
\renewcommand{\tabcolsep}{1pc} 
\begin{tabular}{@{}ccccccc} \hline\hline
Models &  $a^{\sigma}_{1,K}$ & $a^{\sigma}_{2,K}$ & $a^{\sigma}_{3,K}$
& $a^{\sigma}_{4,K}$ & $a^{\sigma}_{5,K}$ & $a^{\sigma}_{6,K}$ \\
\hline
HO  & -0.1501 &  -0.1474 &  0.0198 & -0.0162 & 0.0137 & -0.00036\\
Linear & -0.1262 &  -0.1165 &  0.0031 & -0.0203 & 0.0101 & -0.0031\\
PL    & -0.0218 & -0.0385 & -0.0003 & -0.0090 & 0.0004 & -0.0030 \\
DSE~\cite{DSE2015} & 0.049 & -0.0034 & $\cdots$ & $\cdots$ & $\cdots$ & $\cdots$\\
$\chi$QM~\cite{NK06} & -0.00474 &  -0.1180 &  -0.0030 & -0.0131 &  -0.0007 & -0.0028 \\
\hline\hline
 Models  &  $\la\xi^1\ra^{\sigma}_K$ & $\la\xi^2\ra^{\sigma}_K$ & $\la\xi^3\ra^{\sigma}_K$
& $\la\xi^4\ra^{\sigma}_K$ & $\la\xi^5\ra^{\sigma}_K$ & $\la\xi^6\ra^{\sigma}_K$ \\
\hline
HO & -0.0901 &  0.1495 &  -0.0348 &  0.0503 &  -0.0173 &  0.0227\\
Linear & -0.0757 &  0.1601 &  -0.0319 &  0.0570 &  -0.0169 &  0.0269\\
PL & -0.0131 & 0.1868 & -0.0057 & 0.0760 & -0.0031 & 0.0405 \\
SR~\cite{BBL} & 0.0612 &  0.2022 &  0.0328 &  0.0895 &  0.0221 & $\cdots$  \\
DSE~\cite{DSE2015} & 0.029 & 0.20 &  0.017 &  0.088 &  0.011 & 0.049  \\
$\chi$QM~\cite{NK06} & -0.0028 &  0.1596 &  -0.0018 & 0.0574 &  -0.0012 & $\cdots$ \\
\hline\hline
\end{tabular}
\end{table}
In Table~\ref{t3}, we list the calculated Gegenbauer- and $\xi$- moments of 
the pseudotensor twist-3 kaon
$\phi^\sigma_{3;K}(x)$ 
obtained from the Gaussian wave function with linear and HO potential models and PL wave
function and compare them with other model 
estimates~\cite{BBL,DSE2015,NK06}.
For the kaon case, the odd moments are nonzero due to the flavor SU(3) symmetry breaking effects.
We again note that the sign of Gegenbauer and $\xi$ moments are the same between
our LFQM and $\chi$QM~\cite{NK06} predictions but different from QCDSR~\cite{BBL} and DSE~\cite{DSE2015} predictions.

\section{Summary and Discussion}
\label{sec:IV}

We analyzed the two twist-3 DAs of pion and kaon, i.e. pseudoscalar $\phi^P_{3;M}(x)$ 
and pseudotensor $\phi^\sigma_{3;M}(x)$, within the LFQM. We also
discussed the discrepancy of the asymptotic forms of $\phi^P_{3;\pi}(x)$ between
DSE approach~\cite{DSE2015} and QCD's conformal limit expression~\cite{BF} from the perspective of
dependence of DA on the form of LF trial wave functions, e.g. Gaussian vs. PL wave functions.
While Gaussian wave function satisfies the conformal symmetry in the chiral symmetry limit,
the PL wave function doesn't fulfill the conformal (or scale) invariance in the same limit. 
In order to compute the twist-3 pseudotensor DA $\phi^\sigma_{3;M}(x)$,
we utilized the same manifestly covariant BS model used in~\cite{CJ2015,FB2015,CJ_V14} and then mapped the LF vertex function in the covariant BS model
to the more phenomenologically accessible Gaussian and/or PL wave functions.
Linking the covariant BS model to the standard LFQM, we used the same correspondence (or mapping)
relation given by Eq.~(\ref{Deq:12}) between the two as previously found in~\cite{CJ2015,CJ_V14}.
The remarkable finding in mapping the covariant BS model to the standard LFQM
is that the treacherous points such as the zero-mode contributions and the instantaneous
ones existed in the covariant BS model become absent in the LFQM with the Gaussian
or PL wave function.

Our LFQM descriptions of both twist-3 $\phi^P_{3;\pi}$  and  $\phi^\sigma_{3;\pi}$
satisfy the fundamental constraint(i.e. symmetric form with respect to $x$) anticipated from the isospin symmetry. For the $\phi^P_{3;\pi}(x)$ case, our results with nonzero constituent quark masses show rather convex shapes for both Gaussian and PL
wave functions but they show quite different end point behaviors, i.e. the end points are more enhanced for the PL wave function than
the Gaussian wave function. The difference between the two wave functions are more drastic in the chiral symmetry limit, where the result of Gaussian
wave function  reproduces the  result $\phi^P_{3;\pi}(x)\to 1$  anticipated from the QCD's conformal limit~\cite{BF}
but the result of PL wave function shows the concave shape similar to the result of DSE approach~\cite{DSE2015}. 
This may be understood by the different conformal symmetry behaviors between Gaussian and PL wave functions.
While the authors in~\cite{DSE2015} explained that this difference may come from the mixing effect between the two- and three-particle twist-3 amplitudes, we observe that this difference is linked to the different behaviors of conformal symmetry in the chiral limit of LF trial wave functions. 
For the $\phi^\sigma_{3;\pi}(x)$ case, our results in the chiral symmetry limit, both Gaussian and PL wave functions show very similar shapes each other. Especially, the result from Gaussian wave function reproduces exactly the asymptotic form $6x(1-x)$ anticipated from QCD's conformal limit.
The same chiral-limit behavior was also obtained from the DSE approach~\cite{DSE2015}.
We have now provided the reason why our predictions for the two twist-3 DAs of $\pi$
and chirality-even twist-2 and twist-3 DAs of $\rho$~\cite{CJ_V14} obtained from the Gaussian
wave function in the chiral limit exactly reproduce the forms anticipated from QCD's conformal limit.
For the kaon case, due to the SU(3) flavor-symmetry breaking effect, the twist-3 kaon LF wave functions are distorted in favor of the heavier $s$-quark. The violation of SU(3) flavor symmetry breaking was estimated using Eq.~(\ref{SU3B}) for twist-2 and twist-3 DAs of kaon. The SU(3) flavor symmetry
breakings are $(15, 28,16)\%$ in twist-2 $\phi^A_{2;K}(x)$, 
$(28,38,6)\%$ in twist-3 $\phi^P_{3;K}(x)$, and $(33, 43,5)\%$ in twist-3 $\phi^\sigma_{3;K}(x)$
for (Linear, HO, PL) parameters, respectively. 
In comparison with DSE approach, while our results from the Gaussian wave function are quite different from those of DSE approach, the results from the PL wave function are consistent with those of DSE approach. We may understand these results from the characteristic difference of the conformal symmetry in the chiral symmetry limit of the LF trial wave functions taken for the variational principle computation.

The idea of our LFQM is to provide the nonperturbative wave functions at the momentum scale
consistent with the use of constituent quark mass. The DAs determined from this nonperturbative wave functions can be fed into the QCD evolution equation to provide the shorter distance information of the corresponding hadrons. The DAs obtained without the cutoff should not be regarded as the fully evolved DAs but still be nonperturbative as they
just mean that the cutoff dependence becomes marginal beyond a certain nonperturbative cutoff scale.

\acknowledgments
This work was supported by the Korean Research Foundation
Grant funded by the Korean Government (No.NRF-2014R1A1A2057457).
C.-R. Ji was supported in part by the US Department of Energy
(Grant No. DE-FG02-03ER41260).

\appendix
*\section{Derivation of  twist-3 DAs of a pseudoscalar meson}

Defining $z^\mu = \tau \eta^\mu$ using the lightlike vector $\eta=(1,0,0,-1)$, one can rewrite Eq.~(\ref{Deq:1}) as
\be\label{APP:1}
\la 0|{\bar q}(\tau \eta)i\gamma_5 q(-\tau\eta)|M(P)\ra
= f_M \mu_M \int^1_0 dx e^{i\zeta\tau( P\cdot \eta)} \phi^P_{3;M}(x).
\ee

By integrating  Eq.~(\ref{APP:1}) using the dummy variable $x'$ (and $\zeta'=2x'-1$) with respect to $\tau$ as
\be\label{APP:2}
\int^\infty_{-\infty}  \frac{d\tau}{2\pi} e^{-i\zeta'\tau(P\cdot\eta)} \la 0|{\bar q}(\tau \eta)i\gamma_5 q(-\tau\eta)|M(P)\ra
= f_M \mu_M \int^\infty_{-\infty}  \frac{d\tau}{2\pi}\int^1_0 dx e^{-i(\zeta'-\zeta)\tau( P\cdot \eta)} \phi^P_{3;M}(x),
\ee
and changing the variable $\tau (P\cdot \eta)=T$, we obtain the r.h.s of Eq.~(\ref{APP:2}) as $\frac{ f_M \mu_M}{2(P\cdot\eta)} \phi^P_{3;M}(x')$.
Therefore, the twist-3 $\phi^P_{3;M}(x)$ for pseudoscalar  channel is given by

\be\label{APP:3}
\phi^P_{3;M}(x) = \frac{2 (P\cdot\eta)}{f_M \mu_M}
\int^\infty_{-\infty}  \frac{d\tau}{2\pi} e^{-i\zeta\tau(P\cdot\eta)} \la 0|{\bar q}(\tau \eta)i\gamma_5 q(-\tau\eta)|M(P)\ra.
\ee

Similarly, Eq.~(\ref{Deq:2}) can be rewritten as
\be\label{APP:4}
\la 0|{\bar q}(\tau\eta)\sigma_{\alpha\beta}\gamma_5 q(-\tau\eta)|M(P)\ra
=-\frac{i}{3} f_M \mu_M \tau (P_\alpha \eta_\beta - P_\beta \eta_\alpha)
\int^1_0 dx e^{i\zeta \tau (P\cdot \eta)} \phi^\sigma_{3;M}(x).
\ee

Multiplying $(P^\alpha \eta^\beta - P^\beta \eta^\alpha)$ on both sides of Eq.~(\ref{APP:4}) and using the following identities
$(P^\alpha \eta^\beta - P^\beta \eta^\alpha) (P_\alpha \eta_\beta - P_\beta \eta_\alpha) = -2 (P\cdot\eta)^2$ and
$\sigma_{\alpha\beta} (P^\alpha \eta^\beta - P^\beta \eta^\alpha)=2i (\slash\!\!\!\!P \slash\!\!\!\eta - P\cdot\eta)$, we obtain

\be\label{APP:5}
\la 0|{\bar q}(\tau\eta)(\slash\!\!\!\!P \slash\!\!\!\eta - P\cdot\eta)\gamma_5 q(-\tau\eta)|M(P)\ra
=\frac{1}{3} f_M \mu_M \tau (P\cdot\eta)^2
\int^1_0 dx e^{i\zeta \tau (P\cdot \eta)} \phi^\sigma_{3;M}(x).
\ee

Once again, by integrating  Eq.~(\ref{APP:5}) using the dummy variable $x'$ (and $\zeta'=2x'-1$) with respect to $\tau$ as
\be\label{APP:6}
\int^\infty_{-\infty}  \frac{d\tau}{2\pi} e^{-i\zeta'\tau(P\cdot\eta)} \la 0|{\bar q}(\tau\eta)(\slash\!\!\!\!P \slash\!\!\!\eta - P\cdot\eta)\gamma_5 q(-\tau\eta)|M(P)\ra
=\frac{1}{3} f_M \mu_M (P\cdot\eta)^2
\int^1_0 dx \int^\infty_{-\infty}  \frac{d\tau}{2\pi} \tau e^{-i(\zeta'-\zeta)\tau( P\cdot \eta)} \phi^\sigma_{3;M}(x),
\ee
and changing the variable $2\tau (P\cdot \eta)=T$, we obtain the r.h.s of Eq.~(\ref{APP:6})
as $\frac{i}{12}f_M \mu_M \frac{\partial \phi^\sigma_{3;M}(x')}{\partial x'}$.
Therefore, the twist-3 $\phi^\sigma_{3;M}(x)$ for tensor  channel is obtained by

\be\label{APP:7}
\phi^\sigma_{3;M}(x) = - \frac{12}{f_M \mu_M}
\int^\infty_{-\infty}  \frac{d\tau}{2\pi} \int^x_0 dx'
e^{-i\zeta'\tau(P\cdot\eta)}
\la 0|{\bar q}(\tau\eta) i(\slash\!\!\!\!P \slash\!\!\!\eta - P\cdot\eta)\gamma_5 q(-\tau\eta)|M(P)\ra
\ee

\end{document}